\documentclass[preprint2]{aastex}
\usepackage{psfig}

\include{AAS_defs}

\begin{document}

\title{Accretion Disk Instabilities, CDM models and their role in 
Quasar Evolution}

\author{Evanthia Hatziminaoglou\altaffilmark{1,2}, 
Aneta Siemiginowska\altaffilmark{2} and Martin Elvis\altaffilmark{2}}
 
\altaffiltext{1}{Laboratoire d'Astrophysique de Toulouse, Toulouse, France}
\altaffiltext{2}{Harvard-Smithsonian Center for Astrophysics, Cambridge, USA}

\email{eva@ast.obs-mip.fr,asiemiginowska@head-cfa.harvard.edu,
elvis@head-cfa.harvard.edu}

%%***************************************************************************
%%  ABSTRACT
%%***************************************************************************

\begin{abstract}

We have developed a consistent analytical model to describe the observed
evolution of the quasar luminosity function. Our model combines 
black hole mass distributions based on the
Press - Schechter theory of the structure formation in the Universe with
quasar luminosity functions resulting from
%light curves derived from 
a physics-based emission model that
takes into account the time-dependent phenomena occurring in the
accretion disks. Quasar evolution and CDM models are mutually
constraining, therefore our model gives an estimation of the exponent,
$n$, of the power spectrum, $P(k)$, which is found to be $-1.8 \le n
\le -1.6$.  We were able to reject 
a generally assumed hypothesis of a constant ratio between Dark Matter
Halo and the Black Hole mass, since the observed data could not be
fitted under this assumption.  We found that the relation between the
Dark Matter Halos and Black Hole masses is better described by
$M_{BH}=M_{DMH}^{0.668}$.  This model provides a reasonable fit to the
observed quasar luminosity function at redshifts higher than $\sim$
2.0. We suggest that the disagreement at lower redshift is due to
mergers. 
Based on the agreement at high redshift, we estimated the merger rate at 
lower redshift, and argue that this rate should depend on the redshift, 
like $(1+z)^3$.

\end{abstract}


\section{Introduction}

Since their discovery by Schmidt in the early 60's, quasars have been
an intriguing phenomenon and the mystery surrounding them is still far
from being resolved. Soon after being identified as galactic nuclei,
accretion onto a supermassive black hole became the leading model for
powering these objects (Lynden Bell, 1969), whose luminosities can far
exceed those of galaxies, but whose space density is around two orders
of magnitude lower than that of bright galaxies.  The question of
whether quasar activity was a short-lived phenomenon that occurred
repeatedly during a galaxy's lifetime (Cavaliere \& Padovani, 1989;
Siemiginowska \& Elvis, 1997), or whether quasars never ``wake up''
again after a single, longer-term, activity cycle (Haiman \& Loeb,
1998; Haehnelt, Natarajan \& Rees, 1998), leaving behind massive black
hole remnants, has not yet been resolved.

The quasar population clearly evolves with cosmic time. The quasar
space density above a given luminosity increases by a factor of 
$\sim$ 50 from the present to
$z \sim 2.5$, then stalls and probably decreases to $z \sim 5$ (Hook,
Shaver \& McMahon, 1998). For many years the observed quasar
evolution has been addressed phenomenologically, mostly via Pure
Luminosity Evolution (PLE - Mathez, 1976) or Pure Density Evolution
(PDE - Schmidt, 1968). Both descriptions have just one free adjustable
parameter, but are physically rather abstract. 
Moreover PDE and PLE are observationally indistinguishable when only
a power law is observed, as is common. Instead physically based
models may be constrained by a simple power law, since neither the
slope nor the normalization is now arbitrary. 
An early attempt to
explain the observed quasar space density using a physics-based model
was made by Efstathiou and Rees (1988), who used the Press - Schechter
formalism (Press \& Schechter, 1974) for dark matter halo formation in
order to predict the space density of the more luminous quasars. This
work has been followed by several such attempts, with results that do
not always agree (Cavaliere \& Padovani, 1988; Haehnelt \& Rees, 1993;
Yi, 1996; Siemiginowska \& Elvis, 1997; Haiman
\& Menou, 2000; Haehnelt, Natarajan \& Rees, 1998; 
Salucci et al., 1999). Cavaliere \& Vittorini (1998) suggest that the 
steep rise of the number density of quasars from early epochs to 
a redshift of $\sim$2.5 is due to the newly formed
galactic halos (density evolution) while their fall at lower redshifts
is due to merging processes (luminosity evolution), creating
a link between the phenomenological models and the more robust theory.
Quasars are now a common ingredient of N-body simulations, semi-analytical 
or analytical models, which follow step by step the formation and 
evolution of galaxies within Dark Matter Halos and the merging processes
(e.g. Richstone et al., 1998; Kauffmann \& Haehnelt, 2000; Monaco at al., 
2000). All of the above suggests that galaxy and quasar, both
formation and evolution are closely related, not least because
there is so far no evidence of quasar activity outside galaxies.

Previous work derives the black hole mass distribution from Cold Dark
Matter (CDM) models (Haehnelt \& Rees, 1993; Haehnelt, Natarajan \&
Rees, 1998) or from observations of the local non-active galaxies
(Salucci et al., 1999), but then considers an essentially arbitrary
exponential quasar light curve: a short active phase with luminosities
close to the Eddington limit at the beginning of the black hole's
life, followed by a rapid exponential decay, due to fuel exhaustion
caused by the high accretion rates needed. Such a quasar light curve
is only an approximation of the accretion process and does not include
the physics of the accretion disk which is thought to be formed during
accretion onto a supermassive black hole in quasars (e.g. Lynden-Bell
1969, Shields 1978, Rees 1984). Accretion disks have been studied in
detail in many types of binary systems, where they 
%are never stationary, but 
usually show complex variability.
Both observations and theory point to strong similarities in the
behavior of accretion disks around Galactic black hole X-ray sources
and of those around quasars (Tanaka \& Lewin, 1995; Fiore \& Elvis,
1997; Czerny, Schwarzenberg-Czerny \& Loska, 1999).  These
similarities suggest quite a different quasar light curve. Large
amplitude instabilities typically appear in accretion disks, leading
to repetitive outbursts as large as a factor of $10^4$ on the
timescale of thousand to million years depending on the disk model
(Siemiginowska, Czerny and Kostyunin, 1996, Lin \& Shields 1986).  In
the case of quasar black holes, these variations alone inevitably
create a broad luminosity function (Siemiginowska \& Elvis, 1997).

Unstable accretion leading to episodic outbursts has a valuable
feature.  In our previous paper (Siemiginowska \& Elvis, 1997) quasar
light curves were derived by assuming a continuous fuel supply at a
low constant accretion rate of $\sim 0.04 \dot M_{Edd}$ ( where $\dot
M_{Edd}$ is the critical accretion rate corresponding to the Eddington
luminosity for a given mass).
At this rate a $10^6 M_{\odot}$ black hole accretes some $10^{-3}
M_{\odot}$/yr while a $10^9 M_{\odot}$ black hole accretes
1$M_{\odot}$/yr, so a black hole would need $10^{10}$ years to change
its mass by an order of magnitude.  Consequently the problem of
creating hypermassive (10$^{11} M_{\odot}$ Soltan 1982, Phinney 1983) quasar
remnants does not occur, if the highest initial black hole mass is
10$^9 M_{\odot}$.

In this paper we build on the Siemiginowska \& Elvis (1997) work by
combining the CDM derived black hole mass spectrum with the luminosity
function for a particular mass derived from the time-dependent
accretion disk theory. This results in an end-to-end model for quasar
evolution that is completely physics based, excepting only the ratio
of black hole mass to dark matter halo mass. Since this model links
CDM and quasar evolution, the parameters of CDM constrain quasar
evolution, and the observed quasar evolution constrains CDM
parameters. We examine some of the implications of this model.

\section{Model and assumptions}
\subsection{Accretion onto supermassive black holes}
\label{model}

\begin{figure*}[ht]
\centerline{
\psfig{figure=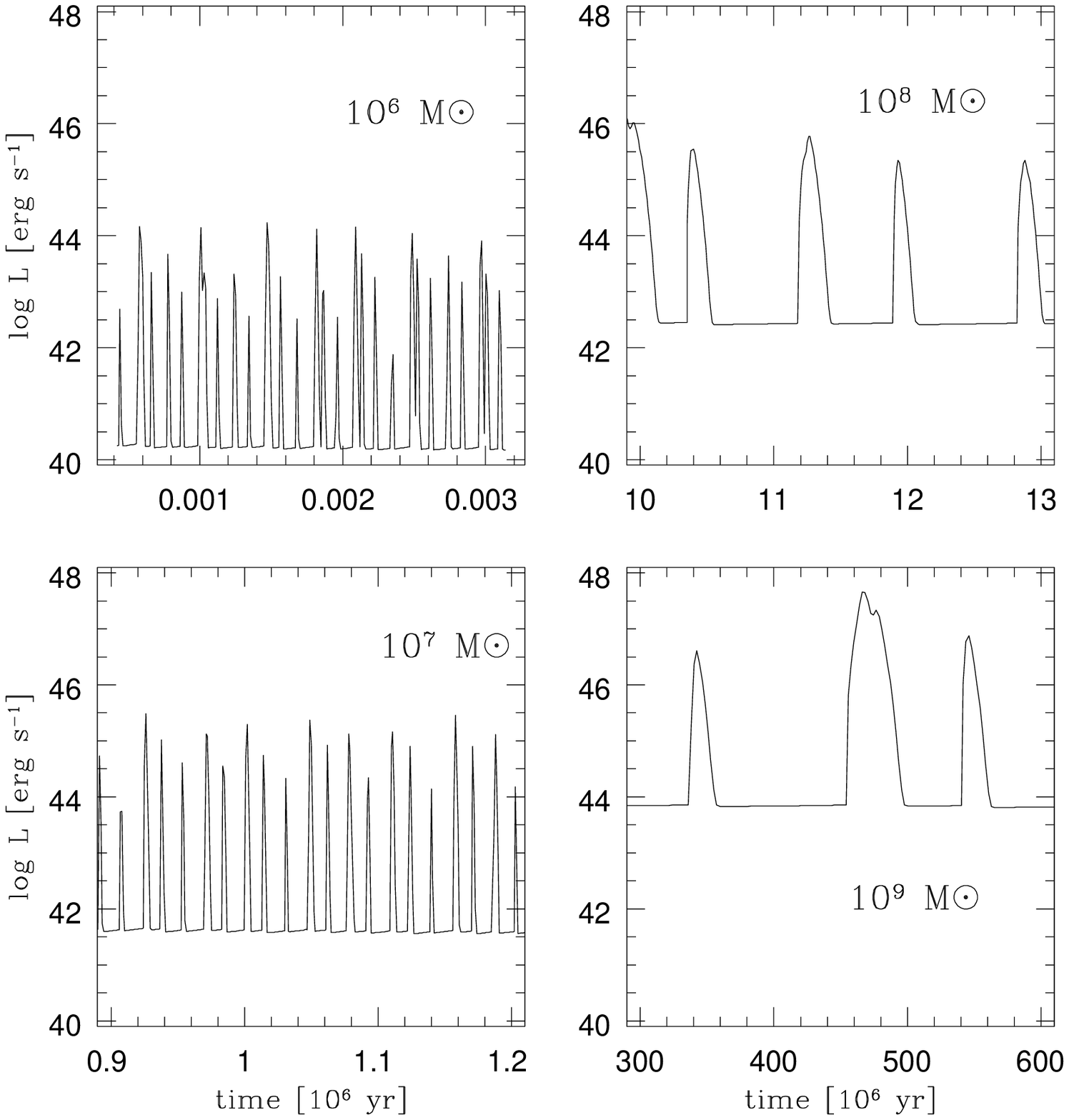,height=12cm,width=12cm}}
\caption{Quasar light curves (accretion disk luminosity vs 
time in 10$^6$ yr) due to thermal-viscous instabilities in the
accretion disk around a central black hole. From top to bottom and
left to right the masses of the black holes are $10^6 M_{\odot}$,
$10^7 M_{\odot}$, $10^8 M_{\odot}$ and $10^9 M_{\odot}$. Accretion
rate at the outer edge of the disk is equal to 0.04$\dot M_{Edd}$
 of critical (Eddington limit) accretion rate.}
\label{lightcurve}
\end{figure*}

We assume that each quasar hosts a supermassive black hole surrounded
by an accretion disk. The disk is the main emission component and
driver of the quasar activity. We calculate a quasar light curve
assuming that quasar emission is due to a non-stationary accretion
disk, thus the quasar luminosity during evolution of accretion disk is
calculated by integrating the disk emission at each epoch.  Our
time-dependent accretion disk model (Siemiginowska et al., 1996) takes
into account thermal-viscous instabilities due to Hydrogen ionization,
which can develop in the disk (Smak, 1982; Lin \& Shields, 1986).
These instabilities are observed in both cataclysmic variables and
Galactic Black Hole Candidates (Smak, 1984; Tuchman, Mineshige \&
Shields, 1990), where they cause factor of 10$^4$ luminosity
outbursts. Similar amplitudes are expected in the disks around
supermassive black holes but the timescales are of order of hundred to
hundred million years (Burderi et al 1998, Siemiginowska, Czerny \&
Kostyunin, 1996; Mineshige \& Shields 1990). As a consequence a single
source can emit within a wide luminosity range depending on the
current state of the disk.  Even if there is a continuous fuel supply
the source will exhibit many active and quiescent states, making
`quasar activity' a recurrent phenomenon occurring numerous times
during a galaxy's lifetime. For example, for a black hole of $10^8
M_{\odot}$, 5 outbursts occur in $\sim 3.2
\times 10^6$ years (figure \ref{lightcurve}).
Siemiginowska \& Elvis (1997) derived the quasar luminosity function
by integrating the time spent by a quasar in each luminosity state and
assuming that this distribution is applied over a whole (uncoordinated)
population with a range of black hole masses.

\begin{figure*}[ht]
\centerline{
\psfig{figure=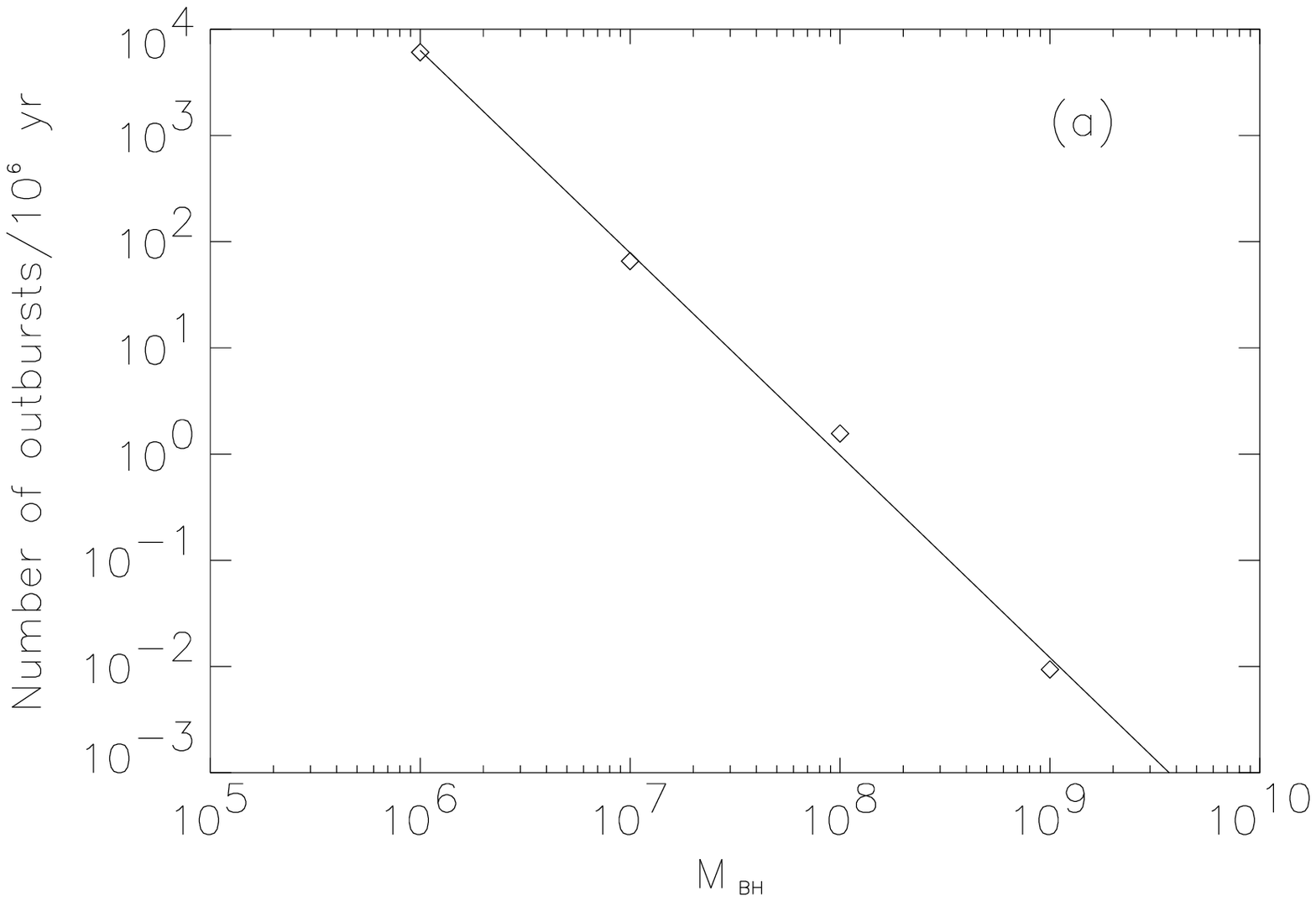,height=8cm,width=8cm}
\psfig{figure=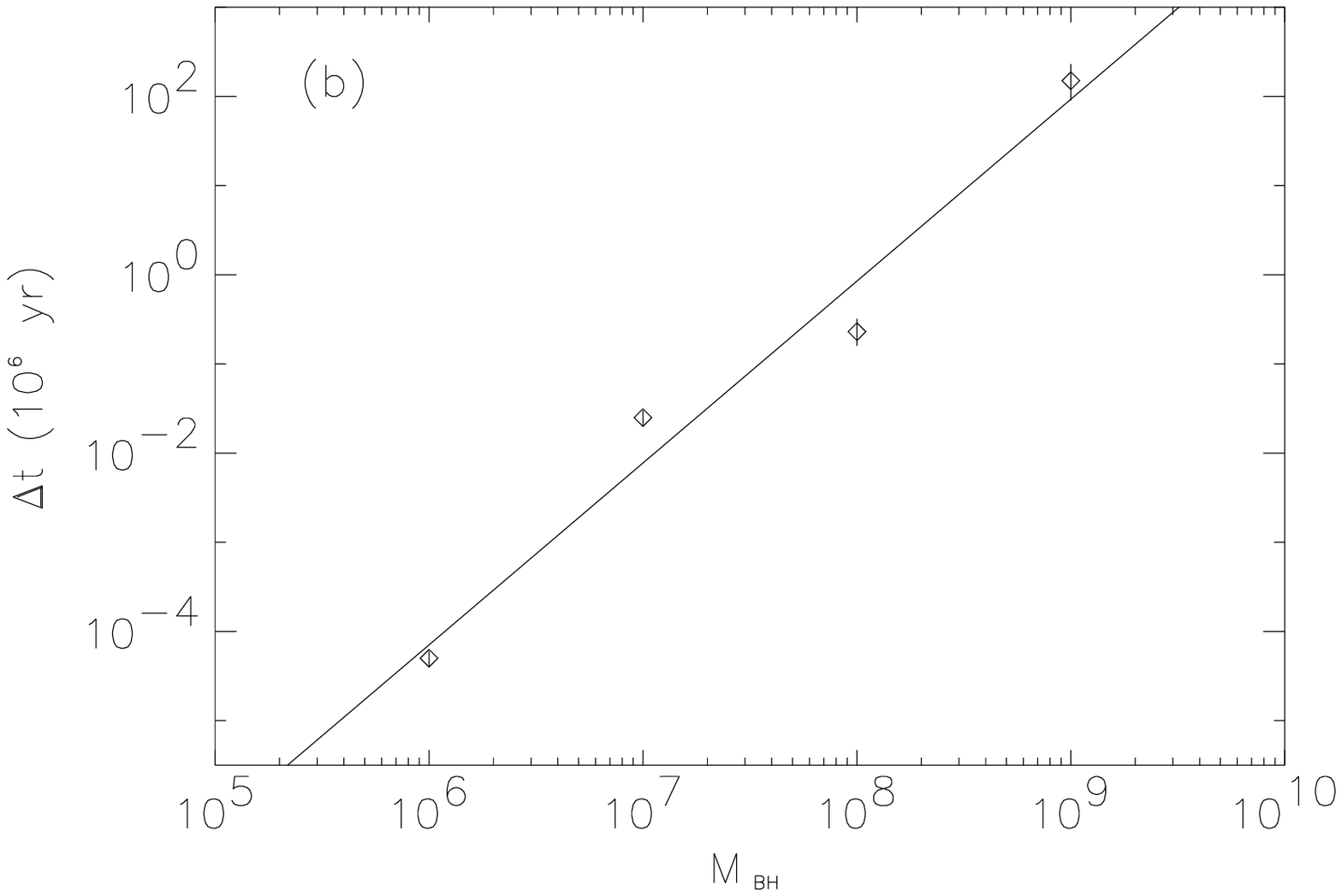,height=8cm,width=8cm}}
\caption{(a) Variation of the number of outbursts
(high luminosity active state) per $10^6$ yr and (b) duration of the
outbursts (high luminosity state) (in $10^6$ yr) versus the black hole
mass. The solid lines illustrate the linear fits.}
\label{mvst}
\end{figure*}

We have checked a simplifying assumption in Siemiginowska \& Elvis
(1997) that the distribution of time in each state is constant for all
black holes masses. In figure \ref{lightcurve} we show the light
curves, due to emission from the accretion disk which exhibits
thermal-viscous instabilities, for black holes with masses $10^6
M_{\odot}$, $10^7 M_{\odot}$, $10^8 M_{\odot}$ and $10^9 M_{\odot}$.
The overall form of the light curves is the same, but for larger
masses, outbursts are rarer than for lower masses, the high luminosity
phases last much longer and the mean luminosity increases too.  Figure
\ref{mvst} quantifies this by showing how the number of outbursts per
$10^6$ yr and the duration of each outburst vary with the black hole
mass.  Both the number of outbursts and their duration depend
quadratically on the black hole mass. 

Note that at $10^6 M_{\odot}$ an outburst occurs every century, which
means that in a sample of 100 AGN we should be able to observe one
transition per year. The Piccinotti et al (1982) 2-10~keV selected
sample has $\sim 20$ non-blazar AGNs, roughly equally divided between
broad and narrow-line objects. In the $\sim 20$ years that these
sources have been studied, one (NGC 7582) has changed from purely
narrow to broad-line status, followed by a considerable change in its
luminosity (Aretxaga et al., 1999).  This is
consistent with the expected statistics. Monitoring of large (N$\sim
1000$) emission-line selected AGN samples for a few years would
strongly test this key premise of the model.

Integration of the light curves over time shows that each source
spends the greater part of its lifetime (around $\sim 70\%$) in a
``quiescent'' state and $\sim 30\%$ in an active state ($L \ge 0.001
L_{Edd}$) with $\sim$ 10$\%$ in a very active state ($L \ge 0.1
L_{Edd}$).  These percentages are almost identical for all black hole
masses, as shown in table \ref{tab}.  Deviations are no larger than
5$\%$ over a range of 4 orders of magnitude, which is
insignificant. The fraction of time a source spends in each activity
state is essentially indeed independent of its mass, as assumed by
Siemiginowska \& Elvis (1997).

\begin{table*}[ht]
\caption{Fraction of time sources of different mass spent in the different 
activity states}
\label{tab}
%\begin{flushleft}
\begin{tabular}{c|ccc}
\hline
\hline\noalign{\smallskip}
M$_{BH} (M_{\odot})$&$L/L_{Edd}<10^{-3}$&$10^{-3} < L/L_{Edd} < 
10^{-1}$&$L/L_{Edd}>10^{-1}$\\
\hline
\hline\noalign{\smallskip}
$10^6$&0.73&0.20&0.07\\
$10^7$&0.75&0.15&0.10\\
$10^8$&0.77&0.14&0.09\\
$10^9$&0.71&0.13&0.15\\
\hline
\hline\noalign{\smallskip}
\end{tabular}
%\end{flushleft}
\end{table*}

\subsection{Dark matter halos and black holes}
\label{massfun}

To derive the black hole mass function, we use the Press - Schechter
(1974) formalism, a simple yet quite adequate approach, which has
become the basis of most formation scenarios of dark matter halos in
CDM models (e.g. White \& Frenk, 1991; Haehnelt \& Rees, 1993).  The
Press - Schechter formalism predicts the mass fraction of the Universe
contained in virialized structures, resulting from a gravitational
hierarchical process (dark matter halos). The mass function of the
dark matter halos (DMHs) can be written as:
\begin{equation}
n(M,z)dM=-\left(\frac{2}{\pi}\right)^{1/2} \frac{\bar{\rho}}{M} \frac{\delta 
_c}{\sigma^2} \frac{d\sigma}{dM}exp\left(-\frac{\delta 
^2_c}{2\sigma^2}\right)dM
\label{PS}
\end{equation}
where $\bar{\rho}$ is the comoving density. This formula gives the
number of dark matter halos of mass $M$ formed up to a redshift $z$.
$\delta _c$, the critical over-density in linear theory for spherical
collapse, has been considered constant, equal to 1.69 (e.g.  Sahni \&
Coles, 1995), although it depends weakly on $\Omega_0$ and $\Lambda$.
The linear theory rms mass density fluctuation in spheres of mass $M$
at redshift $z$ is $$\sigma=\sigma_0\frac{D_z}{D_0}$$ where
$$D_z=(1+3/2\Omega_0)(1+3/2\Omega_0+5/2\Omega_0 z)^{-1}$$ is a
simplified form of the linear growth factor (Peebles, 1980) that
represents the growth of a perturbation in the linear regime.  For the
calculation of $\sigma$ we used the approximation of the present rms
mass density fluctuation in spheres of mass $M$ (e.g. Fan et al.,
1997): $$\sigma_0=\sigma_8\left(\frac{M_8}{M}\right)^{\alpha}$$ where
$M_8=1.19 \times 10^{15}\Omega_0$ is the mean mass in a sphere of
radius 8h$^{-1}Mpc$, $\alpha=(n+3)/6$ where $n$ is the exponent of the
power spectrum $P(k) \propto k^n$, and the corresponding
$\sigma_8=0.67$.  Comparing to other results (Haehnelt, Natarajan \&
Rees, 1998) we verified that the shape of the mass function for masses
greater than $10^8 M_{\odot}$ is not affected by this approximation.

\begin{figure*}[ht]
\centerline{
\psfig{figure=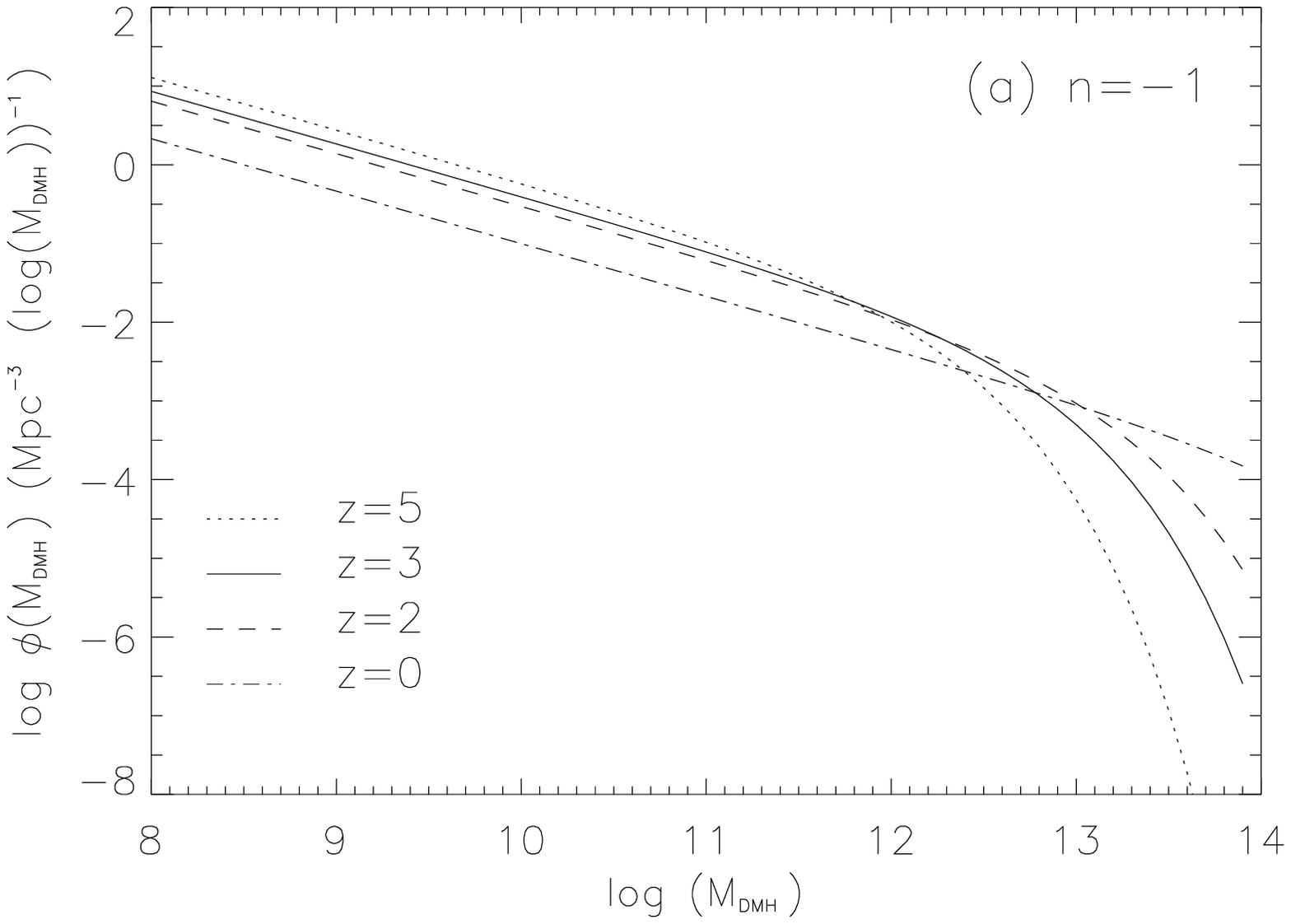,width=8cm,height=8cm}
\psfig{figure=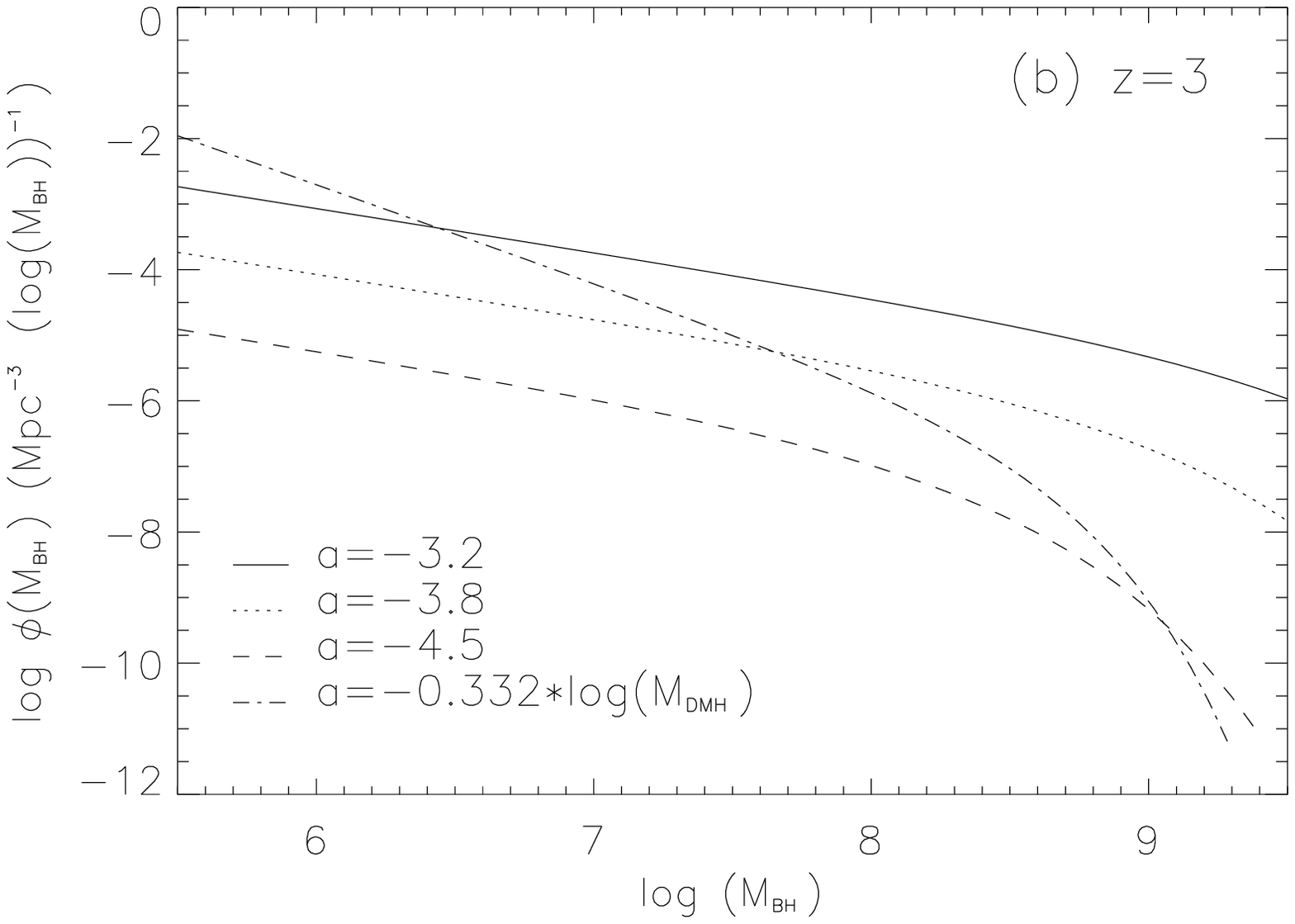,width=8cm,height=8cm}}
\caption{(a) Press - Schechter 
DMH mass function for z=5 (dotted line), 3
(solid line), 2 (dashed line) and 0 (dash-dotted line)
for the cosmological model q$_0$=0.5 and for $n=-1$. (b) Black hole
mass functions for $z=3$. The solid line corresponds to $\alpha$=--3.2, the
dotted line to $\alpha$=--3.8, the dashed line to $\alpha$=--4.5 and the
dash-dotted line to $\alpha=-0.332 \times log(M_{DMH})$.}
\label{massfunps}
\end{figure*}

The DMH mass function for the standard CDM model, for $n=-1$ and for
various redshifts between 5 and 0 is illustrated in figure \ref
{massfunps}a. A single cosmological model (q$_0$=0.5) will be used henceforth.

Equation \ref{PS} refers to halos harboring both elliptical and spiral
galaxies. Quasars are found in both elliptical and spiral galaxies
(McLeod et al., 1995; McLeod \& Katris, 1995; Taylor et al., 1996;
Bahcall et al., 1997; Boyce et al., 1998). Narrow lined (type 2) AGN
can also reside in both early type galaxies and in spirals
(e.g. Salucci et al., 1999).  We assume that the distinction between
quasars and Seyferts is purely one of luminosity (we also include both
radio-loud and radio-quiet AGN in term `quasar'), the latter occurring
in spiral galaxies, since we consider that they are both the
manifestation of the same basic phenomenon.  The black hole mass
function can be directly derived from the dark matter halo mass
function.  We assume that in each newly formed dark matter halo a
black hole is ``instantly'' created, i.e. that all galaxies harbor a
black hole. This hypothesis has been supported by several observations
(e.g. Keel, 1983; Ho et al., 1997; Magorrian et al., 1998; Richstone
et al., 1998; Ford et al., 1998).  Haiman \& Loeb (1998) assumed that
black hole formation is restricted to halos whose masses are bigger
than $10^8 ((1+z)/10)^{-3/2} M_{\odot}$ with the mass of the black
hole being a constant fraction of the dark matter halo mass:
$M_{BH}=10^{\alpha} M_{DMH}$, where $\alpha$ =--3.2 for all black hole
masses.  We adopt this mass relation as a baseline, but examine the
possibilities of varying $\alpha$ between --4.5 and --3.2.  The
resulting black hole mass functions are shown in figure
\ref{massfunps}b.  Values close to --3.2 are consistent with the
$M_{BH}/M_{bulge}=0.006$ of Magorrian et al. (1998), while values
close to --4.5 are consistent with the $M_{BH}/M_{bulge}=10^{-3.5}$,
given by Wandel (1999), where $M_{bulge}$ denotes the mass of the galactic
bulge.  We also consider the possibility of an
$\alpha$ that is a function of the DMH mass, $\beta \times
log(M_{DMH})$, which is equivalent to a non-linear relation between the BH
and DMH masses, $M_{BH}=M_{DMH}^{1+\beta}$. 
The dash-dotted line in figure \ref{massfunps}b traces
the black hole mass function for $\beta=-0.332$, a value chosen by
fitting the observations, as explained in the next section.

\section{Deriving the luminosity function}
\label{qsolumfun}

The luminosity function is the product of the fraction of time each
source of a given mass spends in each luminosity (activity state) and
their space density (Siemiginowska \& Elvis, 1997).  More formally,
the luminosity function of a population of quasars is the result of
the convolution of their mass function with the light curve, L(t), of
an individual source of each mass:

\begin{equation}
\Phi(L,z)=\int \Phi(L,z,M)n(z,M)dM,
\label{LF}
\end{equation}

\noindent
where $n(z,M)$ is the mass function and $\Phi(L,z,M)$ is the 
fraction of objects that contribute to a given luminosity at a given 
redshift (equivalent to $f_{on}$ in equation \ref{HMLF}). 

This is equivalent to the Haiman \& Menou (2000) suggestion that the
observed quasar luminosity function $dN_{obs}(z,M)/dL$, can be
directly compared with the Press - Schechter DMH mass function,
$dn(z,M)/dM$, to derive the luminosity of each halo of mass $M$:

\begin{equation}
\frac{d\Phi _{obs}(z,M_{BH})}{dL} \frac{dL}{dM}=f_{on}\frac{dn(z,M)}{dM},
\label{HMLF}
\end{equation}

\noindent
where $f_{on}$ is the quasar ``duty cycle'' defined as the fraction of
black holes that are active at a given redshift, $z$.  Note that
equation \ref{LF} is the integrated form of equation \ref{HMLF}, if
the halo mass function is replaced by the black hole mass function.

\begin{figure*}[ht]
\centerline{
\psfig{figure=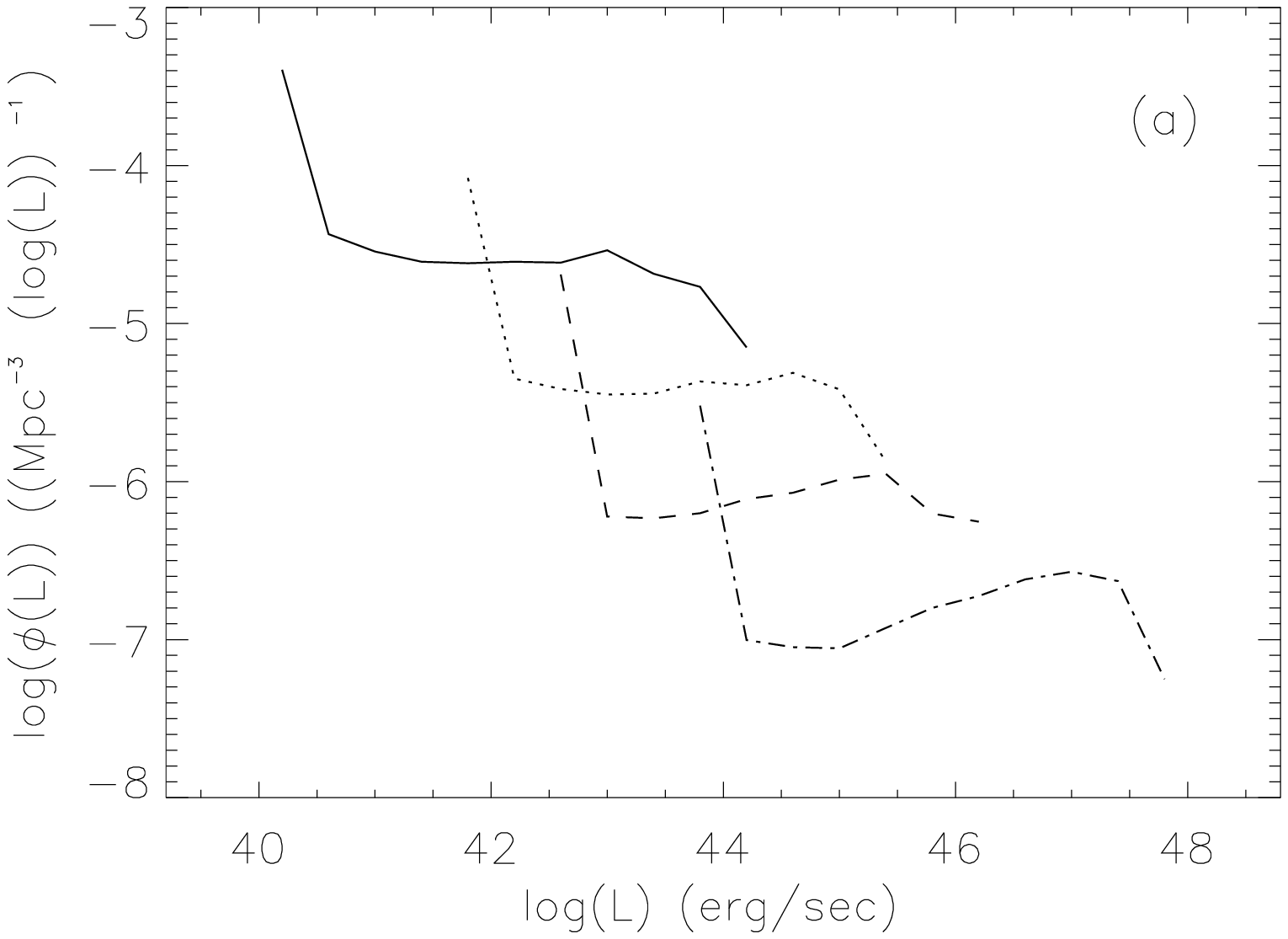,width=8cm,height=8cm}
\psfig{figure=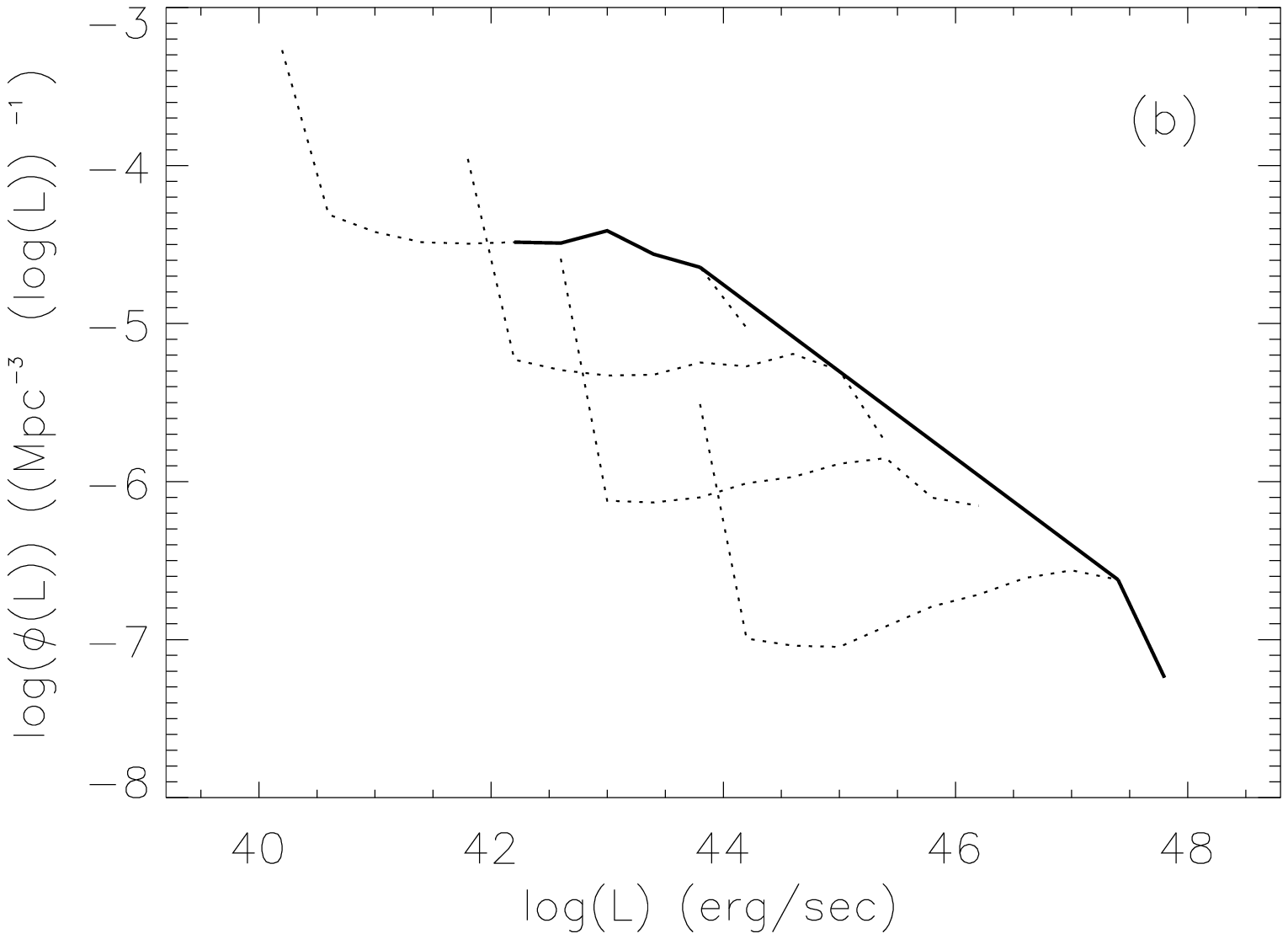,width=8cm,height=8cm}}
\caption{(a) Contribution of the different black hole mass sources to the luminosity function at
$z=2$, for black hole masses ranging from $10^6 M_{\odot}$ (solid line) 
to $10^9 M_{\odot}$
(dash-dotted). 
$\alpha$ was constant and equal to --3.2 for all black hole
masses. (b) The slope defined by the sources at high luminosity state
gives roughly the slope of the luminosity function. In the case of an
upper (here 10$^9 M_{\odot}$) or lower (here 10$^6 M_{\odot}$) mass
limit, the luminosity function will follow the shape of the
corresponding light curves, creating two breaks.}
\label{convol}
\end{figure*}

By integrating the light curves over timescales that allow many
outbursts to occur, we calculated the time a source spends in the
different activity states for a range of black hole masses. This is
equivalent to calculating, for a certain redshift, $z$, the fraction
of sources of a given mass, $M$, in a given activity state, $L$,
$\Phi(L,z,M)$.  All the above are illustrated in figure \ref{convol}a,
where we show the contribution of sources with black hole masses from
$10^6 M_{\odot}$ to $10^9 M_{\odot}$, to the overall luminosity
function, for a Press - Schechter derived black hole mass 
function at a redshift of 2.  The sources with black
hole masses within the abovementioned range have luminosities within
the observed limits (10$^{41} erg/sec \le L \le 10^{48} erg/sec$).
Masses have been converted to luminosities through the light curves
and $\alpha$ was held constant, $M_{BH}=10^{-3.2} M_{DMH}$, for all
black hole masses.

\begin{figure*}[ht]
\centerline{
\psfig{figure=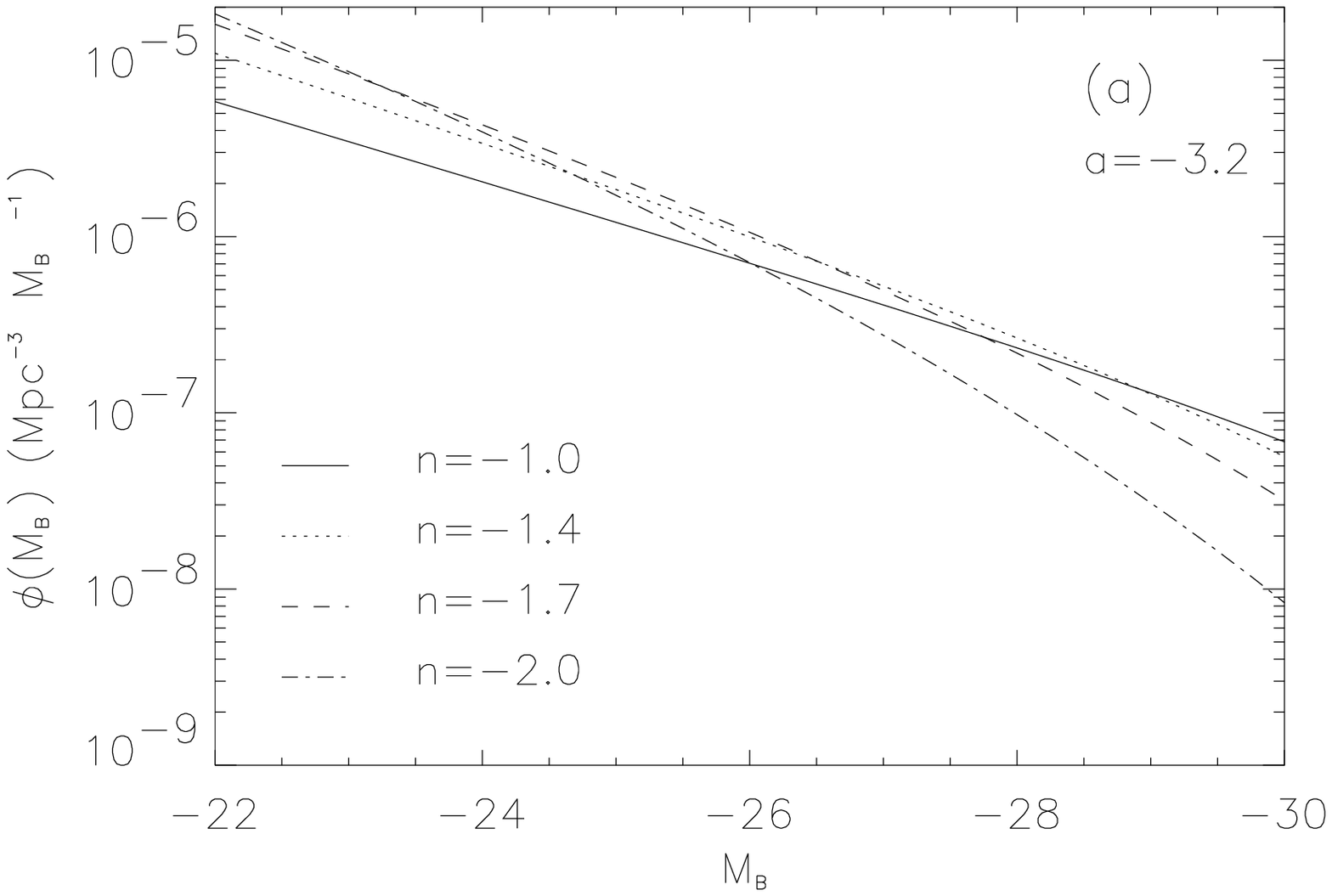,width=8cm,height=8cm}
\psfig{figure=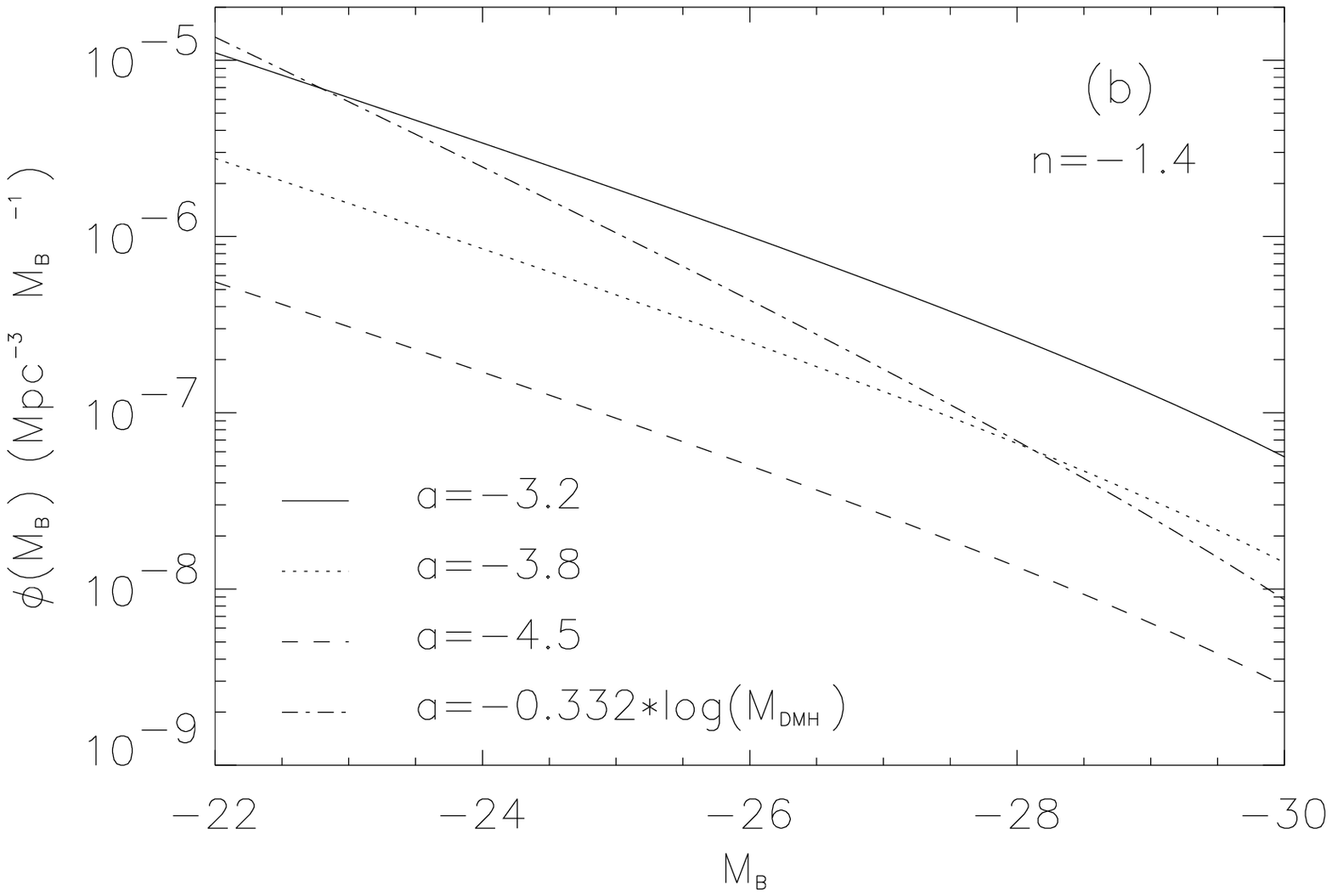,width=8cm,height=8cm}}
\caption{Predicted luminosity function for $z=2$. 
(a) $n$ varies while
$\alpha$ is kept constant, (b) $\alpha$ varies while $n$ is constant.}
\label{LFz2}
\end{figure*}

Note that in each case, a low luminosity state for a high mass source,
is a high luminosity state for a lower mass source, and vice
versa. Figure
\ref{convol}b shows that at
each redshift, the slope defined by the high activity sources gives
roughly the shape and slope of the luminosity function.
In the case of an upper or lower mass limit,
the luminosity function will follow the shape of the corresponding
light curves, creating two breaks: one towards the lower and one
towards the higher luminosities.  Figure
\ref{LFz2} illustrates the luminosity function of the form
$\Phi(M_B)dM_B$, for different values of the parameter $n$
(Fig. \ref{LFz2}a) and the parameter $\alpha$ (Fig. \ref{LFz2}b), including
black hole masses 10$^6 M_{\odot} \le M_{BH} \le 10^9M_{\odot}$,
after converting the luminosities into absolute magnitudes,
$M_B$.  The shape of the
luminosity function and its evolution depend on these two parameters.
Our discussion is focussed on the effect of varying these parameters.

\subsection{Effects of changing CDM Power Spectrum Slope ($n$)}
\label{pwl}

\begin{figure*}[ht]
\centerline{
\psfig{figure=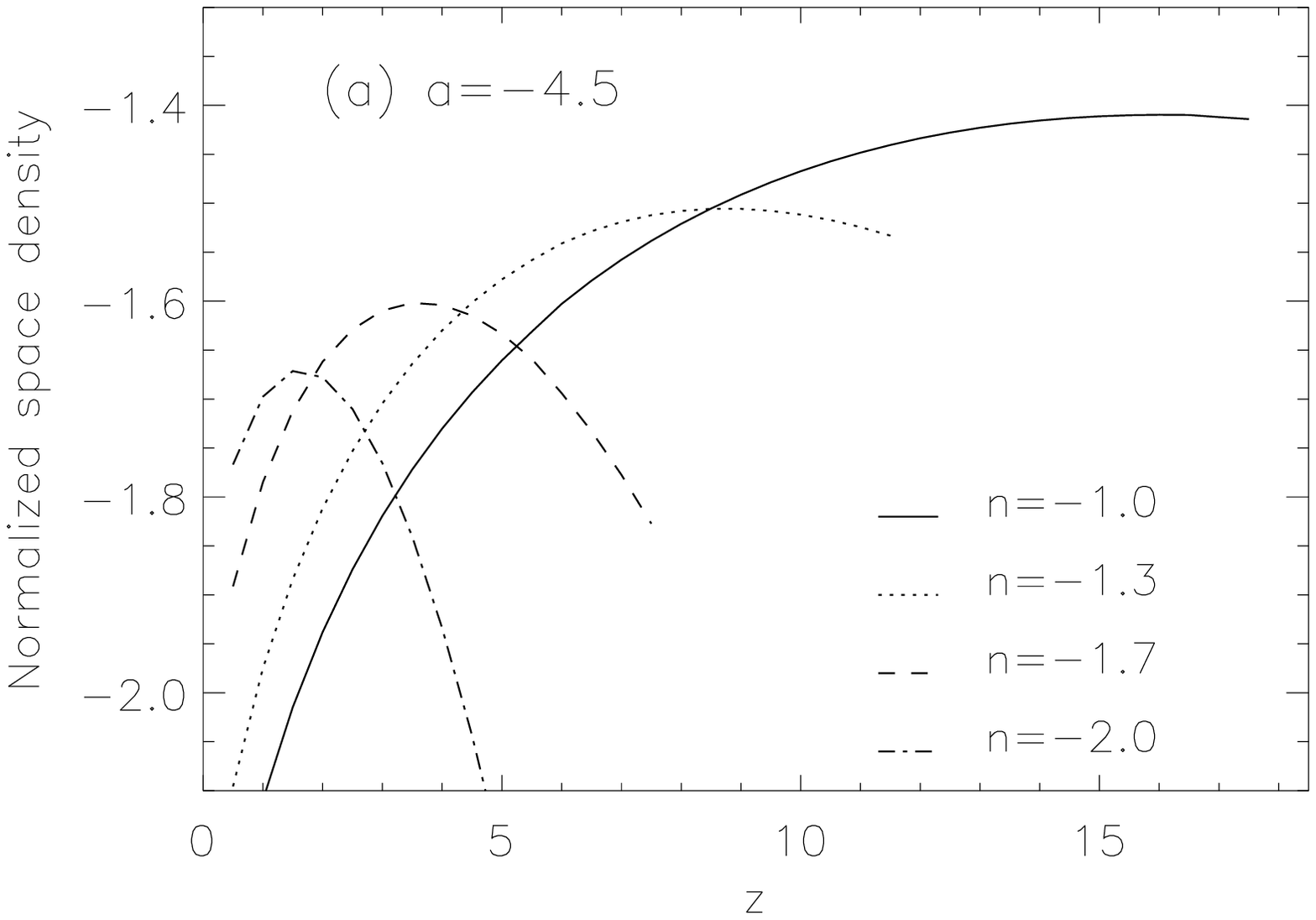,width=8cm,height=8cm}
\psfig{figure=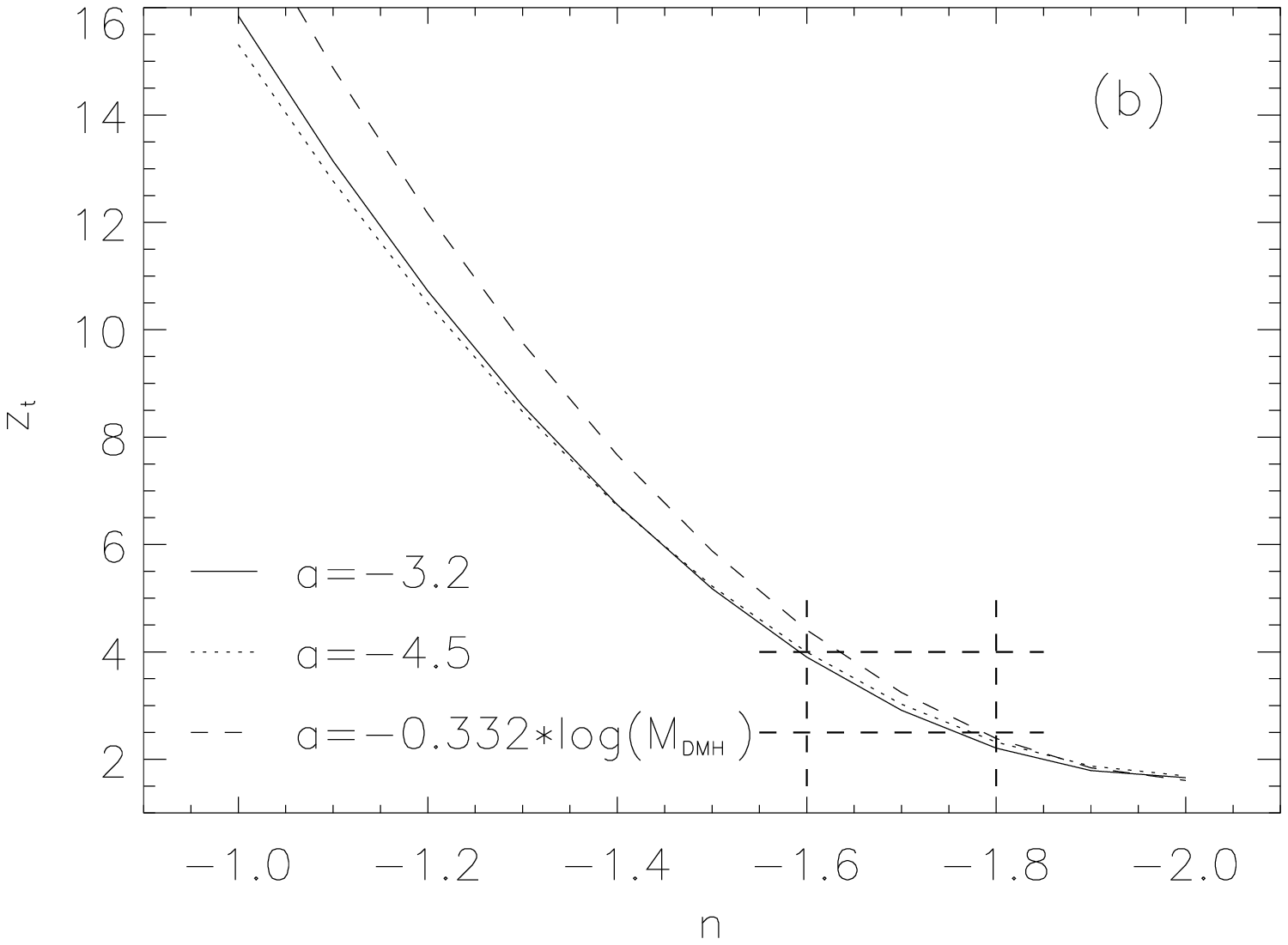,height=8cm,width=8cm}}
\caption{(a) Predicted logarithmic
normalized quasar space density for
$n$=--1.0 (solid line), $n$=--1.3 (dotted line), $n$=--1.7
(dashed line) and $n$=--2.0 (dash-dotted line) for $\alpha$=--4.5.
(b) Variation of $z_t$ with the power spectrum exponent,
$n$, for $\alpha$=--3.2 (solid line), $\alpha$=--4.5 (dotted line) and
$\alpha=-0.332 \times log(M_{DMH})$. The relation is
parabolic and almost independent of $\alpha$. The dashed lines mark the
accepted values for $z_t$, deduced from observations and the constraints
they impose on the values of $n$.}
\label{zvsn}
\end{figure*}

Here we examine the effect of changing the values of the exponent $n$
of the CDM power spectrum.  For a CDM-type spectrum $P(k) \propto
k^n$, $n$ takes values in the interval $-2 < n < -1$ on scales of
$\sim 8h^{-1}$ Mpc (e.g. Fan et al., 1997).  For many years, the
galaxy correlation function data were consistent with $n=-1.2$
(Peebles, 1980; Davis \& Peebles, 1983). However, new surveys show
that the shape of the power spectrum has a more complicated form
and/or that the values of $n$ vary between small and large scales
(e.g. Lin et al., 1996; Hermit et al., 1996; Le Fevre et al., 1996;
Tadros et al., 1999).  We do not attempt to introduce a more complex
form for the power spectrum.  However, we do find that varying the
exponent, $n$ produces strong effects on the quasar luminosity
function.  Observations show that the quasar space density increases
as $\Phi(z)=\Phi_0 (1+z)^3$ from the current epoch up to a turnover
redshift, $z_t$, of 2-3, where it stops or possibly drops (e.g.  Hook
et al., 1998). Our models naturally produce a $z_t$ by integrating
over the whole redshift range the predicted luminosity function, for
sources with absolute magnitudes brighter than $M_B=-26$.  We find
that the $z_t$ predicted by our models is a parabolic function of
$n$. Figure \ref{zvsn}a shows the logarithmic 
normalized space density of quasars
brighter than $M_B=-26$ as a function of $z$, for 4 different values
of $n$ (with $\alpha=-4.5$). The peak of the distribution, $z_t$,
changes with $n$: the steeper the power spectrum the lower the
$z_t$. Figure \ref{zvsn}b shows the relation between $z_t$ and $n$,
which is roughly parabolic.  Moreover, this relation is almost
independent of the value of $\alpha$ making this a robust result,
since $\alpha$ is the only physically undetermined parameter in the
model.

The observed $z_t$ is at least as high as 2.5 (Hook et al., 1998) and
may be as high as $\sim$ 4 (Miyaji et al., 2000), but quasar space
density declines rapidly for $ z > 4$, as suggested by both optical
and radio observations (Kennefick et al., 1997; Hook et al., 1998) and
theoretical models (e.g. Cavaliere \& Vittorini, 1998, give a $z_t$
around 3.5). This puts a strong constraint on $n$, which can only take
the values between --1.6 and --1.8 (figure \ref{zvsn}b). These values
are in excellent agreement with the values given by Blanchard et al.,
1999, derived from the galaxy clusters. We restrict the rest of our
study to this range of $n$.

\subsection{Dependence of Halo to black hole mass ratio ($\alpha$) on mass.} 

In the next two sections we discuss how the parameter $\alpha$ affects
the luminosity function and we consider two possibilities: (1)
$\alpha$ is constant for all masses; (2) $\alpha$ is related to the
mass of the DMH.

\subsubsection{Constant ($\alpha$)}
\label{acst}

\begin{figure}[ht]
\centerline{
\psfig{figure=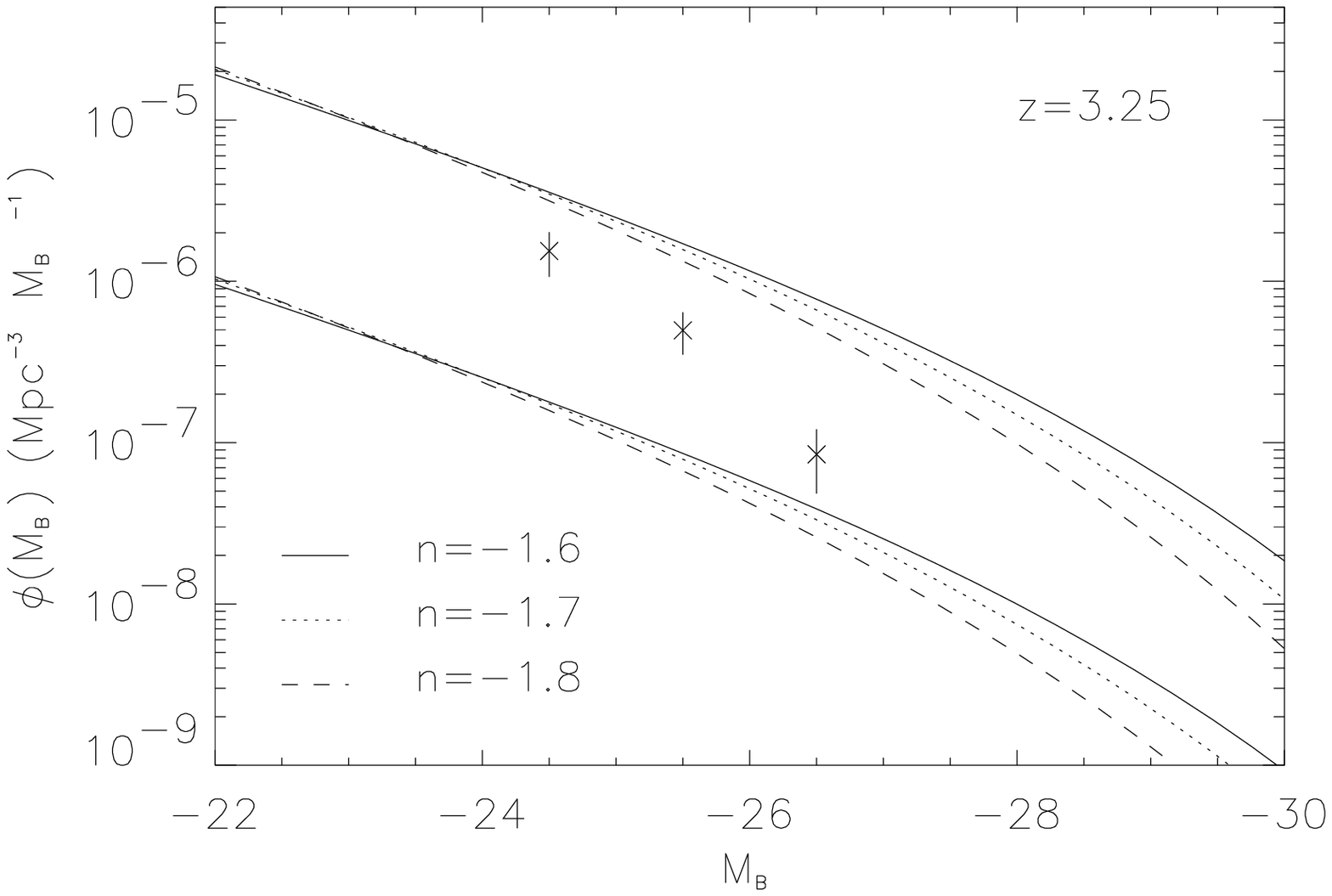,width=8cm,height=8cm}}
\caption{Predicted luminosity function at $z=3.25$
for different values of $n$ for constant $\alpha$.  Upper set of
curves: $\alpha=-3.2$, lower set of curves: $\alpha=-4.5$.  The
symbols show the observed luminosity function from Warren et al
(1994).}
\label{csta}
\end{figure}

Most theoretical models have assumed for simplicity that the black
hole mass is a constant fraction of the dark matter halo mass,
independent of the masses involved (e.g. Haiman \& Loeb, 1998; Haiman
\& Menou, 2000). Observationally the results of Magorrian et al (1998)
suggest that this is roughly obeyed.  Therefore we first examine
this possibility and compare our predictions with observations, both
for low and for high redshifts.  Figure \ref{csta} illustrates the
predicted luminosity function in the case of $\alpha=-3.2$ and
$\alpha=-4.5$ for different values of $n$ ranging from --1.6 to --1.8,
at $z=3.25$. The symbols are the observed luminosity function from
Warren et al. (1994).

In both cases, the slope of the predicted luminosity function is
different and much smoother than the observed luminosity function and
no break is present.  The observed luminosity function cannot be
fitted with the constant $\alpha$ model.  For a high value of $\alpha$
(--3.2) the luminosity function is over-estimated at high
luminosities, while for a low value ($\alpha=-4.5$) the numbers of
faint quasars are under-predicted. The same tendencies are seen at
all redshifts. This suggests that $\alpha$ cannot be constant, but
may be a function of the masses involved.

\subsubsection{$\alpha$ dependent on the $M_{DMH}$}
\label{avar}

\begin{figure*}[ht]
\centerline{
\psfig{figure=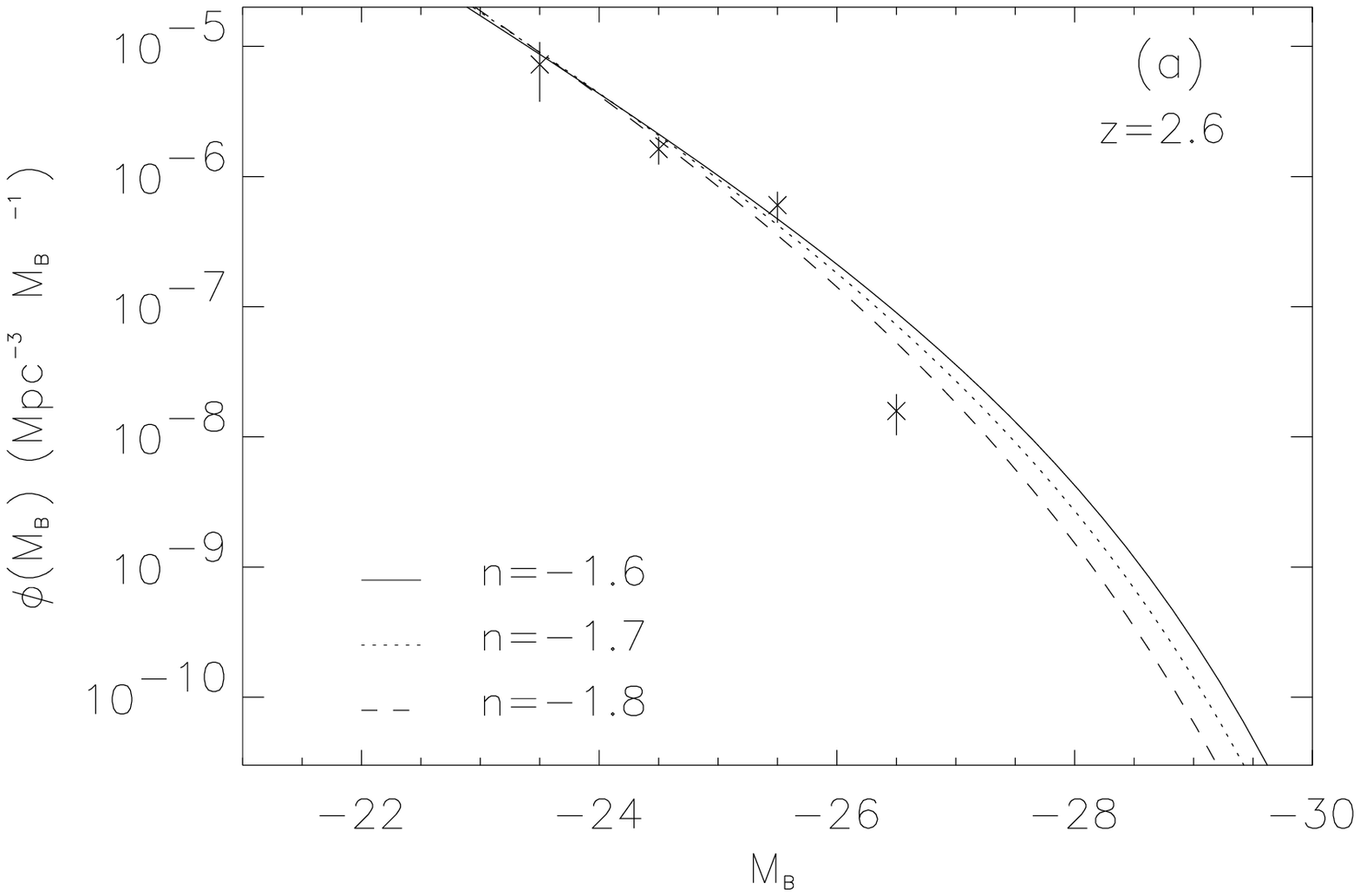,width=8cm,height=8cm}
\psfig{figure=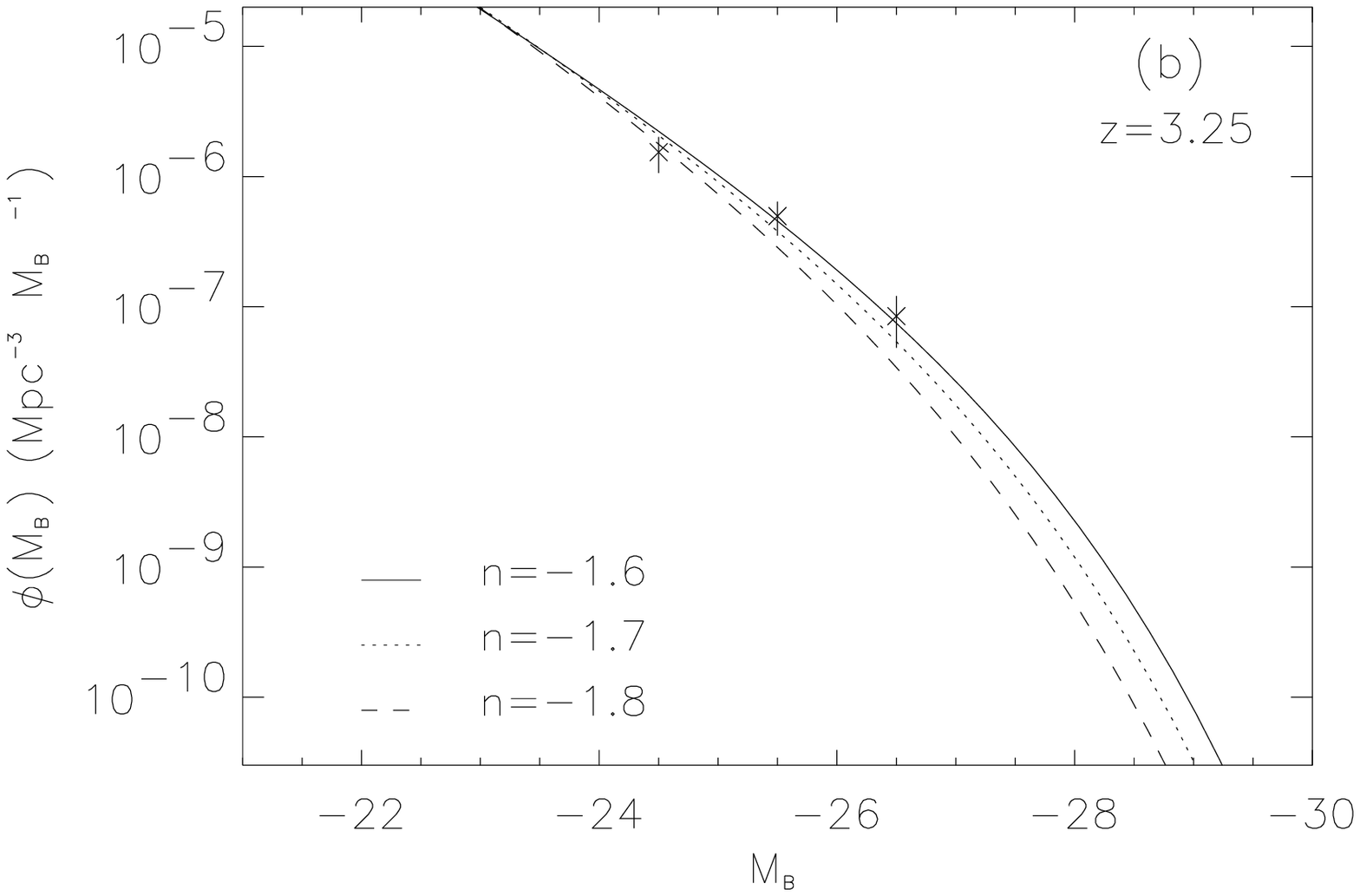,width=8cm,height=8cm}}
\caption{Predicted luminosity function 
for $z=2.6$ and 3.25, for $n$=--1.6
(solid line), --1.7 (dotted line) and --1.8 (dashed line). Symbols show
the observed luminosity function of Warren et al. (1994).}
\label{LF2z}
\end{figure*}

There is no a priori reason that the black hole mass and the dark
matter halo mass should have constant ratio, so in this section we
discuss the effects of varying $\alpha$.  We assume that $\alpha$
depends on the DMH's mass with the relation $\alpha =
\beta \times log(M_{DMH})$, or, in other words, that $M_{BH}=
M_{DMH}^{1+\beta}$.  In order to obtain the best value of a
parameter $\beta$ we fit this function to the observed luminosity
function from Warren et al. (1994) at redshifts 2.6 and 3.25, shown in
figure \ref{LF2z}.  We found that $\beta = -0.332^{+0.04}_{-0.016}$
(or 1+$\beta = 0.668$)
gives the best fit, which is the value used below.

The black hole
masses are now defined from the relation: $M_{BH}=M_{DMH}^{0.668}$. 
This relation gives $\alpha$=--3 for $M_{BH}=10^6
M_{\odot}$ and $\alpha$=--4.4 for $M_{BH}=10^9 M_{\odot}$, which are
in agreement with the observed values (Magorrian et al., 1998 and
Wandel, 1999).

Figure \ref {LF2z} illustrates the predicted luminosity function for
$z=2.6$ and $z=3.25$ for 3 values of $n$ from --1.6 to --1.8.  The
symbols show the Warren et al. (1994) data. The fit is much better
than for any constant value of $\alpha$. The bright end of our predicted 
luminosity functions
at $z=2.6$ seems to be over-predicted, when compared to observations
(figure \ref{LF2z}a). Alternatively, the observed bright end may have 
been under-estimated, as suggested by recent surveys, in particular the 
Hamburg/ESO survey (Wisotzki et al., 1996, Koehler et al, 1997 and Wisotzki,
2000). We conclude that $\alpha$ must be a function of the DMH in order to
properly describe the most recent observations.

Our physical model leads to changes in the luminosity function
equivalent to either PLE or PDE, up to a redshift of $\sim$ 4, because
its form is close to being a simple power law.

\subsection{Comparison with observations - discussion}
\label{obs}

\begin{figure*}[ht]
\centerline{
\psfig{figure=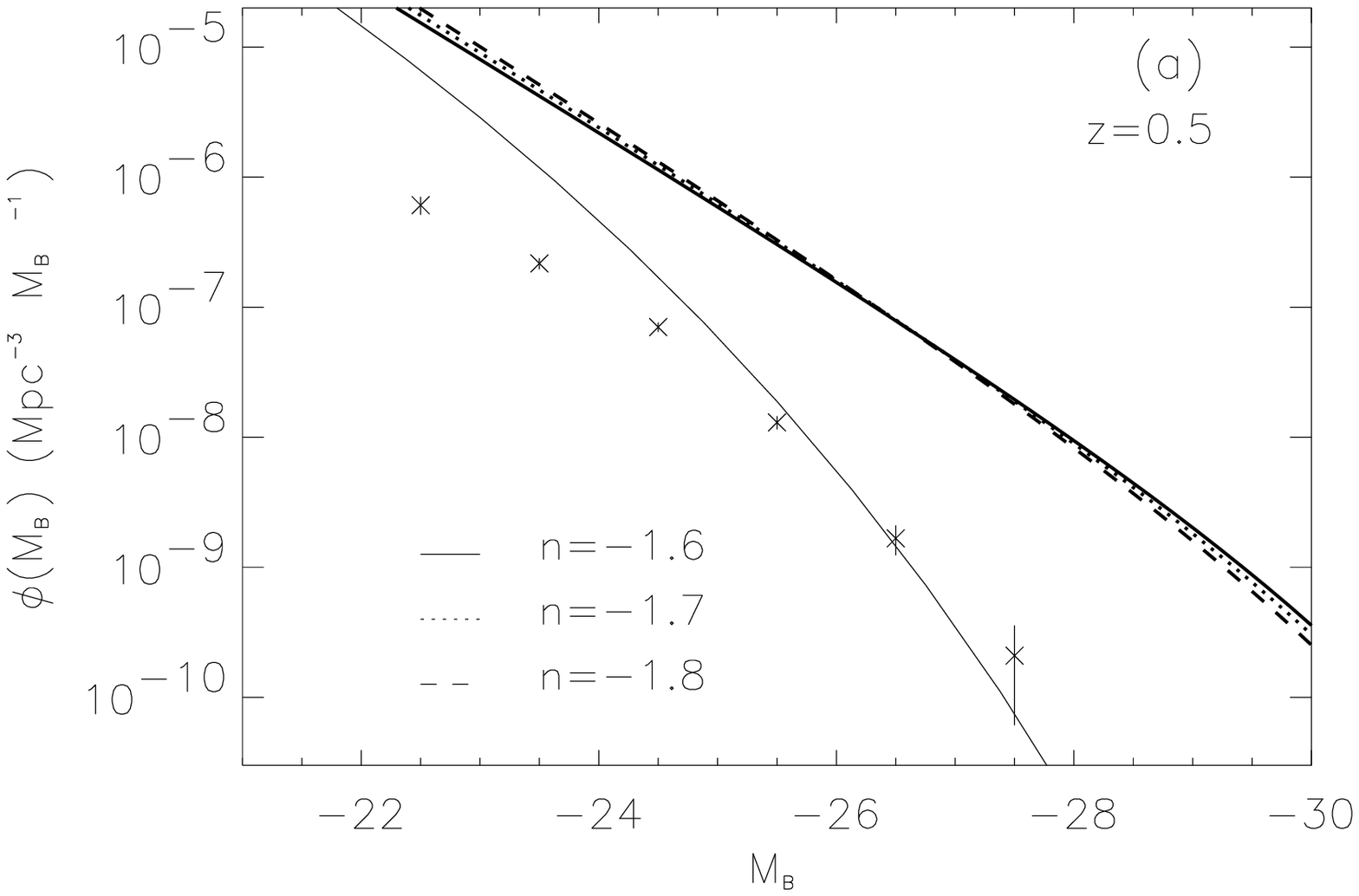,width=7.5cm,height=7.5cm}
\psfig{figure=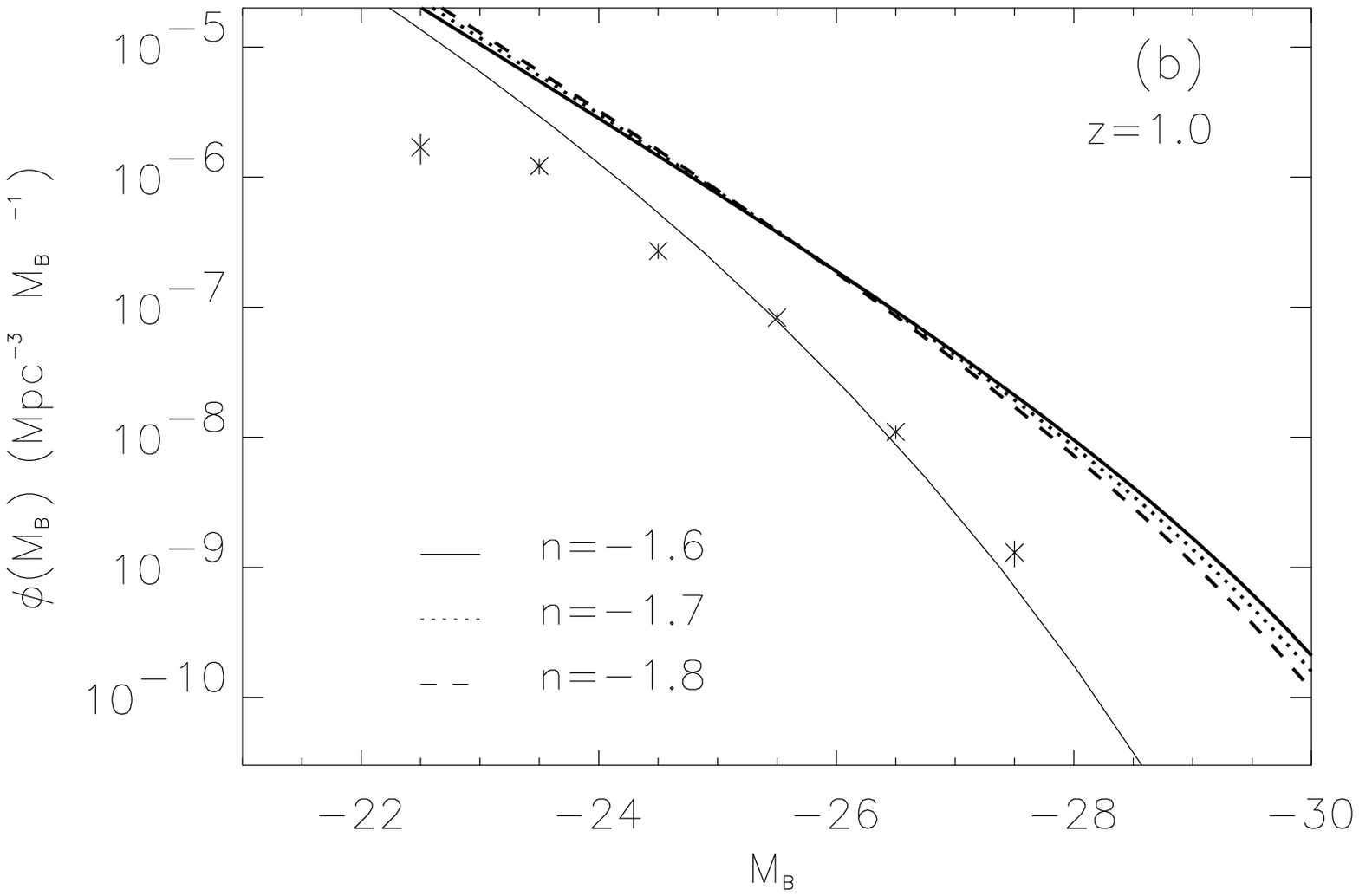,width=7.5cm,height=7.5cm}}
\centerline{
\psfig{figure=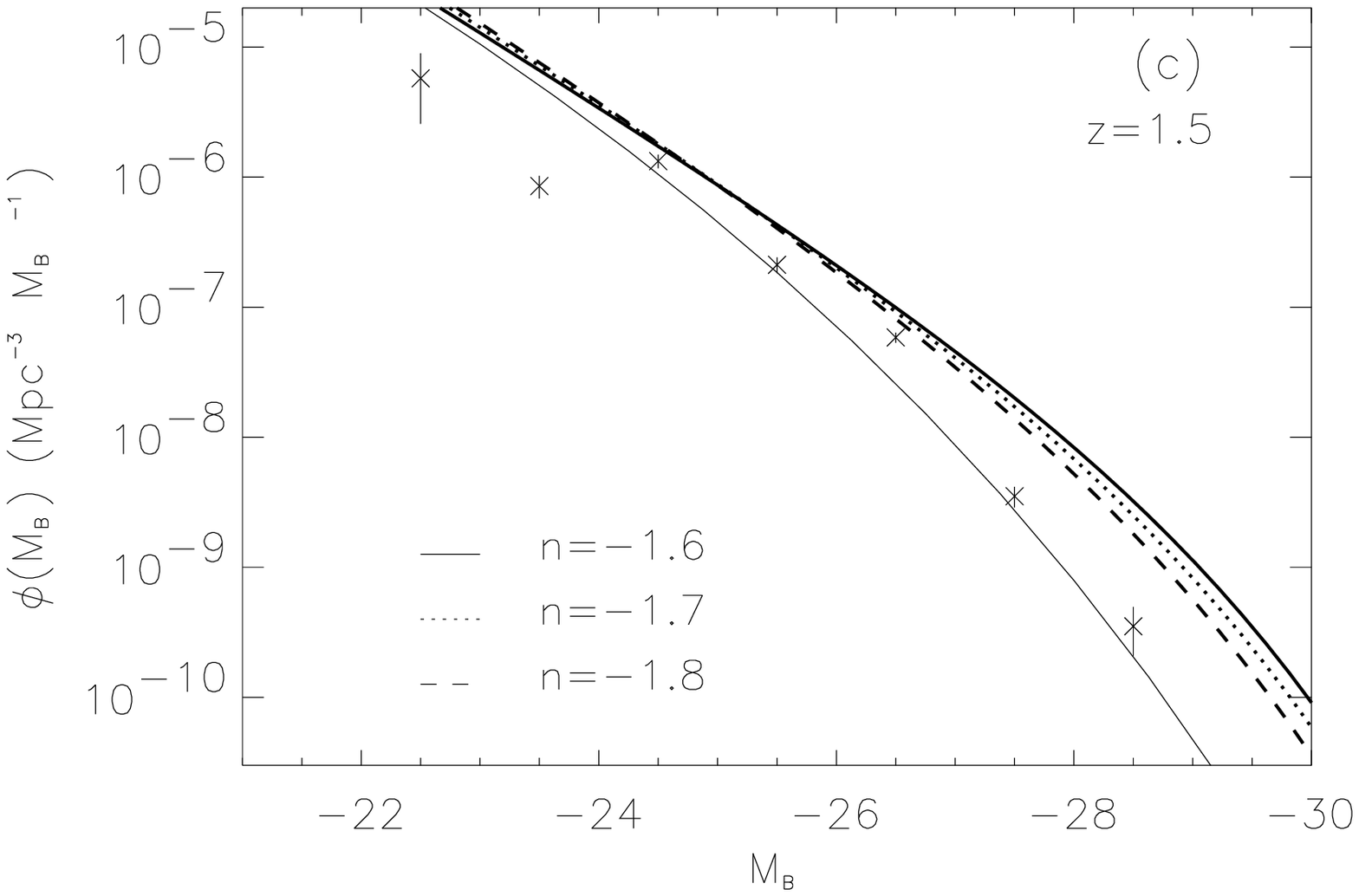,width=7.5cm,height=7.5cm}
\psfig{figure=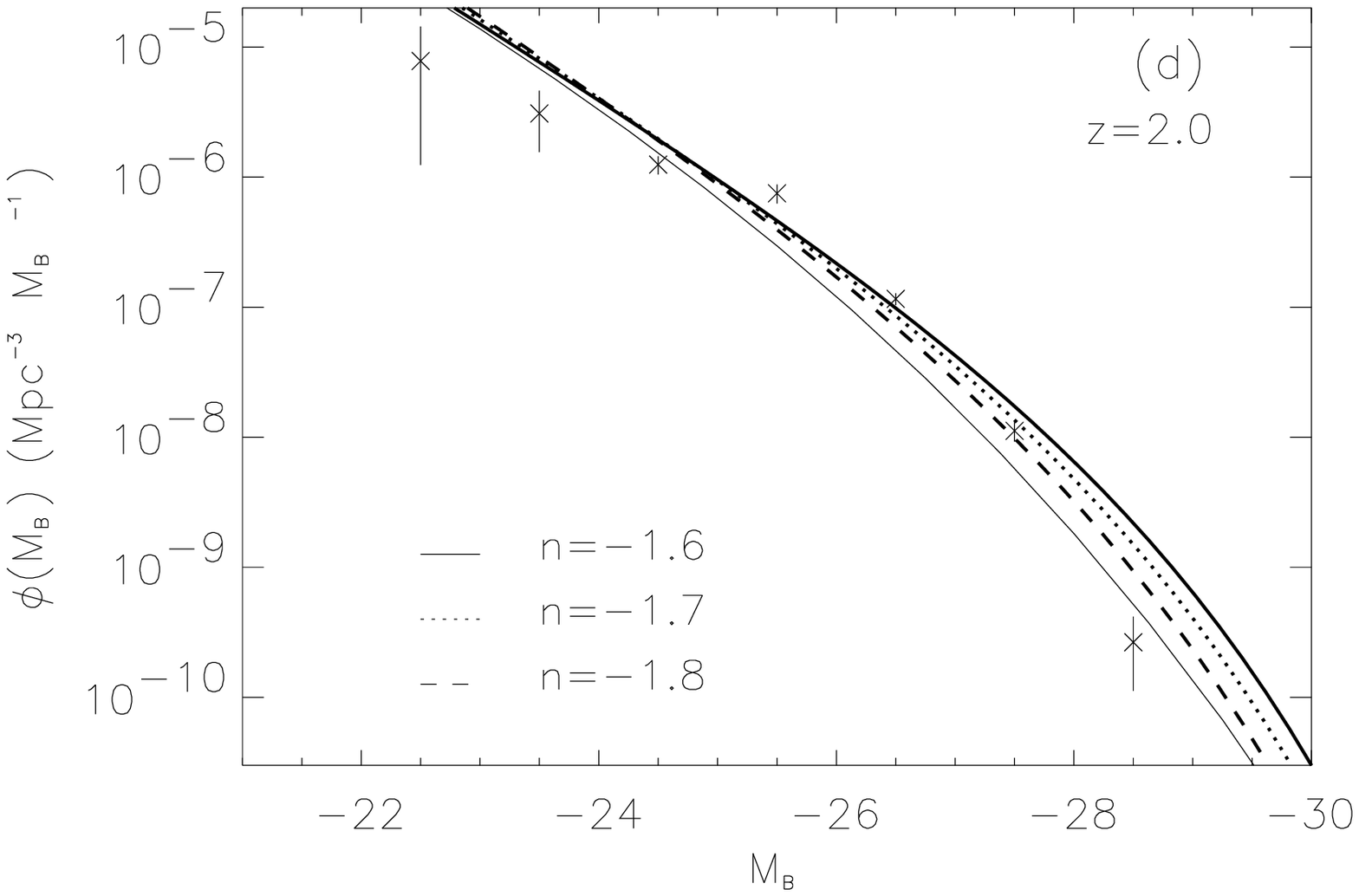,width=7.5cm,height=7.5cm}}
\caption{The predicted 
quasars luminosity function for --1.8 $< n <$ --1.6
at (a) $z=0.5$, (b) $z=1.0$, (c) $z=1.5$ and (d) $z=2.0$ (thick lines). 
The symbols
represent the observed luminosity function for the samples AAT (Boyle
et al., 1991), LBQS (Hewett et al., 1995), HES (Wisotzki et al., 2000)
and CFRS (Le Fevre et al., 1995) (Wisotzki, private communication). 
The thin solid line represents the single power law
LF of Pei (1995).}
\label{LFB}
\end{figure*}

The predicted evolution fits the data quite well at high redshifts,
e.g $z>2.0$.  For redshifts below $\sim 2$, however, the predicted 
luminosity function, in terms of PLE with $L=L_=(1+z)^{\gamma_p}$ evolves as
$\gamma_p=0.5$. This disagrees with the observed luminosity function 
issued from the
following samples: AAT (Boyle et al., 1991), HES (Wisotzki et al.,
2000), LBQS (Hewett et al., 1995) and the CFRS quasars (Le Fevre et
al., 1995), for redshifts 0.5 to 2.0 (figure \ref{LFB}), that evolves
much more strongly with, $\gamma_o \sim 3.5$, although the slopes of the
two (observed and theoretical) luminosity finctions are the same, as 
seen in figure \ref{LFB}. In this figure, the symbols show the observed
luminosity function issued from the abovementionned samples while the dashed,
dotted and thick solid lines illustrate the predicted luminosity function
for $n=-1.8$, $n=-1.7$ and $n=-1.6$, respectively.

For many
years, the best established form of the observed quasar luminosity
function was a broken power law, first given for optically selected
quasars with redshifts $0.3 \le z \le 2.2$ by Boyle et al. (1988),
confirmed by La Franca \& Cristiani (1997) for other optical surveys
and by Page et al. (1996) for X-ray selected AGN.  However, recent
surveys, in several wavelengths, show that this assumption was
probably due to selection effects.  For $z > 2.2$ (as Ly${\alpha}$
shifts from the U to the B-band) multicolor selection techniques
result in biased samples and favor the redshift regions at which
quasar colors make them distinguishable from stars.  Giallongo and
Vagnetti (1992) argued that the observed break was due to the
uncertainties on the K-correction due to the dispersion in quasar
optical slopes. This seems to be confirmed by ROSAT AGN data (Miyaji
et al., 2000). More recent surveys, which account for this effect,
such as the Hamburg/ESO Quasar Survey (which uses emission lines in
addition to colors, Wisotzki et al., 1996, Koehler et al, 1997 and
Wisotzki, 2000) show that the number of bright objects was
underestimated and that the combined {\it local} luminosity function
for Seyferts and quasars could be fitted with a single unbroken power
law, as argued by Giallongo and Vagnetti (1992).  Pei (1995) also
suggests that a single power law of the form $L^{1/4}$ could fit the
data as well as a broken power law. This single power law is represented
by the thin solid line in figure \ref{LFB}. In terms of luminosity, galaxies
with central black hole masses of the order of $10^6 M_{\odot}$ will
be seen as Seyfert galaxies, while the higher masses will result in
quasars (figure \ref{convol}).

Even if the observations are altered somewhat additional physical
effects are likely to be important, since we derived the mass function
using only the Press - Schechter DMH mass function.  For example
galaxy mergers could significantly modify the assumed mass function
(e.g. Haiman \& Loeb, 1998; Kauffmann \& Haehnelt, 2000).  Mergers are
rare in the early Universe but start becoming important at a redshift
around 2, an epoch of massive galaxy formation (Kats et al., 1992;
Kontorovich et al., 1992).  Only mergers that result in the formation
of a single object will cause a decrease of the space density of the
black holes. The black holes, due to the dynamical friction, will soon
be found in the center of the newly formed object and merge
(e.g. Binney \& Tremaine, 1994), thus influencing the quasar space
density. If this is the primary cause of the low $z$ disagreement of
our model with observation then the difference of the observed to
predicted evolution rates, $\gamma_o - \gamma_p$, provides a crude
estimate of the merger rate, $\gamma_m$, for redshifts between 0.5 and
2, i.e. $\gamma_m = 3$. Hence the merger rate can be expressed as a
function of the redshift: $\sim (1+z)^3$. Interestingly this value is
in agreement with several other results based on observations: Zepf \&
Koo (1989) studied a complete sample of 20 close pairs of faint (B
$\le$ 22) galaxies and found that the frequency of interacting
galaxies increase as $(1+z)^{4.0 \pm 2.5}$. Burky et al. (1994) used
four fields observed with the Hubble Space Telescope and found that
the galaxy merger rate increases with redshift as $(1+z)^{2.5 \pm
0.5}$. Patton et al. (1997) used a sample of 545 field galaxies and
estimated that the merger rate changes with redshift as $(1+z)^{2.8
\pm 0.9}$.

Note that less extreme galaxy interactions, where the objects remain
independent entities afterwards will not influence our results, if we
assume that whatever happens in the vicinity of an object will not
influence its accretion rate and therefore its light curve.  In
reality, such interactions are likely to affect the accretion rate and
will be taken into account in the future work.

The disagreement between the observed and the predicted luminosity
function at $z \le 2$ may also be due to the role of obscured (type
II) AGN.  The estimates of their number density are not yet accurate.
Recent work (Salucci et al., 1999) estimates the local mass function
of the dormant black holes within elliptical galaxies, assuming that
the nuclear activity is a single short event. Salucci et al., however,
do not take merger effects into account, introducing uncertainties in
the calculated massive dark object mass function.

We used a black hole mass distribution with black holes
ranging from $\sim$ 10$^6 M_{\odot}$ to 10$^9
M_{\odot}$. According to our model, accretion onto such black holes
can produce luminosities within the observed range $ 10^{41} \le L \le
10^{48}$ erg/sec ($10^{-4} \le L/L_{Edd} \le 1$). If we allow for a
higher black hole mass cut-off at $\sim 10^7 M_{\odot}$ we could
obtain the break in the luminosity function. However, there is 
evidence for black holes with lower masses
(e.g. Reid et al., 1999, Sag A$^* \sim 2.6 \times 10^6 M_{\odot}$).
Therefore a break cannot be imposing a simple cut-off mass.

Our model gives a consistent physical connection between high and low
luminosity sources and does not predict a break in the luminosity
function for the black hole mass distribution within the
observed mass range.

The model presented here can be extended and used for the study of the
UV-background and the re-ionization history of the Universe
(Siemiginowska et al., 2000, in preparation).  Issues such as whether
quasars are entirely responsible for the re-ionization of the Universe
or whether Population III stars are required (see Madau 1999),  the
redshift at which it occurred, whether it is homogeneous or
inhomogeneous or its effects on the primordial power spectrum, can be
examined. The results are likely to be different from previous
attempts because the emission spectrum derived from a non-stationary
accretion disk is different from the one predicted by a stationary
model at the UV wavelength range (Siemiginowska et al., 1997).

\section{Conclusions}

In this paper we presented a theoretical model for calculating the
quasar luminosity function. This model, unlike previous ones, takes
into account the physical processes occurring in the accretion disk
around a black hole. Black holes are assumed to form instantly within
all dark matter halos, with masses related to the halo mass as:
$M_{BH}=M_{DMH}^{0.668}$.  

We first restricted the range for the two free parameters of the
model: the power spectrum exponent, $n$ and the ratio of black hole
mass to DMH mass, $\alpha$.  For any reasonable $\alpha$ value, ($-4.5
< \alpha < -3.2$), $n$ has to be within the narrow range --1.6 to
--1.8 in order to match the observed redshift of the peak of quasar
evolution.  We compared the predicted luminosity function to the data
and found that a non-linear relation between the BH and DMH masses 
($M_{BH}=M_{DMH}^{1+\beta}$) gives a much better
approximation than a linear relation ($M_{BH}=10^{\alpha}M_{DMH}$).
Our model predicts
no break in the luminosity function, assuming quasars and Seyfert
galaxies to be objects of the same kind but of different
luminosities. Observationally this model is then equivalent to both
PLE and PDE, but physically has quite different implications.

No assumptions have been made on the merger processes, although they
impact the quasar evolution.  Mergers that result in the black hole
fusion (and consequently in a decrease of the black hole space
density) will also influence the luminosity function at redshifts
lower than $\sim 2$, when mergers become important. We estimated that
the merger rate should depend on redshift, $(1+z)^3$, consistent with
direct measurements.

Built into our model is the idea that quasar-like activity is a
recurrent phenomenon. Practically all galaxies go through that state
many times during their lives. Recent observations support the idea
that black holes reside in all galaxies.

%\tableofcontents
\acknowledgments

This work was supported by NASA grants NAG5-3391, NAG5-4808,
NAG5-6078, and NAS8-39073.

\end{document}